\documentclass[11pt]{article}
\usepackage[margin=1in]{geometry}
\usepackage{xspace}
\usepackage{url}
\usepackage{hyperref}
\usepackage{tikz}

\title{NP-Completeness and Physical Zero-Knowledge Proof\\ of Hotaru Beam}

\author{Taisei Otsuji, Peter Fulla, Takuro Fukunaga\\ \\ Faculty of Science and Engineering, Chuo University, Tokyo, Japan\\ 
\texttt{otsuji0024@gmail.com}, \texttt{\{fulla,fukunaga\}@ise.chuo-u.ac.jp}}

\date{}

\newtheorem{theorem}{Theorem}
\usetikzlibrary{patterns}

\newcommand{\club}{\raisebox{-1.1mm}{\tikz \node[card] {\large $\clubsuit$};}\xspace}
\newcommand{\heart}{\raisebox{-1.1mm}{\tikz \node[card]{\large $\heartsuit$};}\xspace}
\newcommand{\spade}{\raisebox{-1.1mm}{\tikz \node[card]{\large $\spadesuit$};}\xspace}
\newcommand{\diam}{\raisebox{-1.1mm}{\tikz \node[card]{\large $\diamondsuit$};}\xspace}
\newcommand{\blank}{\raisebox{-1.1mm}{\tikz \node[card]{};}\xspace}
\newcommand{\numcard}[1]{\raisebox{-1.1mm}{\tikz \node[card]{#1};}\xspace}
\newcommand{\bbinary}[1]{\raisebox{-1.1mm}{\tikz \node[card]{#1};}\xspace}
\newcommand{\bclub}{\raisebox{-1.1mm}{\tikz \node[bcard]{\large $\clubsuit$};}\xspace}
\newcommand{\bheart}{\raisebox{-1.1mm}{\tikz \node[bcard]{\large $\heartsuit$};}\xspace}
\newcommand{\bspade}{\raisebox{-1.1mm}{\tikz \node[bcard]{\large $\spadesuit$};}\xspace}
\newcommand{\bdiam}{\raisebox{-1.1mm}{\tikz \node[bcard]{\large$\diamondsuit$};}\xspace}
\newcommand{\bblank}{\raisebox{-1.1mm}{\tikz \node[bcard]{};}\xspace}
\newcommand{\bnumcard}[1]{\raisebox{-1.1mm}{\tikz \node[bcard]{#1};}\xspace}

\tikzset{card/.style={draw, inner sep=0pt, outer sep=0pt,minimum width = 1em, minimum height=1.1em}}
\tikzset{bcard/.style={draw, inner sep=0pt, outer sep=0pt,minimum width = 1em, minimum height=1.1em,color=red,pattern=dots,pattern color=red}}
\tikzset{hotaru/.style={draw, line width=1.5pt, fill=white,circle,inner sep=0pt, outer sep=0pt,minimum width =15pt}}
\tikzset{dot/.style={fill=black,circle, inner sep=0pt, minimum width=5pt}}
\tikzset{beam/.style={very thick}}
\tikzset{backg/.style={gray, thin}}

\begin{document}

\maketitle

\begin{abstract}
Hotaru Beam is a logic puzzle which objective is to connect circles placed on a
grid by drawing only lines with specified starting points and numbers of bends.
A zero-knowledge proof is a communication protocol that allows one player to
persuade the other that they are in possession of a certain piece of information
without actually revealing it. We show that Hotaru Beam is NP-complete and
present a physical zero-knowledge proof (i.e. implementable using physical
items) for proving that one knows a solution to the puzzle.
\end{abstract}

\section{Introduction}

A \emph{zero-knowledge proof} (ZKP) is a communication protocol between two players, a
\emph{prover} and a \emph{verifier}. Using a ZKP, the prover can persuade the
verifier that they possess a piece of information without actually revealing it.
The concept of ZKPs was introduced by Goldwasser, Micali, and
Rackoff~\cite{GoldwasserMR85}. Since then, it has been regarded as an important
concept in both cryptography and complexity theory.

A \emph{physical ZKP} refers to a ZKP that can be carried out by humans using everyday items. 
Gradwohl\ et\ al.\ \cite{GradwohlNPR09} 
noted  that physical ZKPs are easily understandable even for non-experts, which enhances the implementability and reliability of the protocols. 
The development of physical ZKPs for puzzles has been actively studied in recent decades. Examples of these puzzles include
Sudoku~\cite{GradwohlNPR09}, Nurikabe and Hitori~\cite{robert2021interactive}, Slitherlink~\cite{0001MMSS19}, 
Numberlink~\cite{RuangwisesI21b}, and Bridges~\cite{RuangwisesI21}.
Goldreich~et~al.\ \cite{GoldreichMW91}
demonstrated that a ZKP can be constructed for any problem in NP. 
While this finding can be applied to obtain ZKPs for these puzzles, designing ZKPs tailored to specific puzzles is advantageous due to their potential for simplicity and efficiency compared to those given by Goldreich et al.

\emph{Hotaru Beam} is a logic puzzle created by the Japanese puzzle company
Nikoli (see Wikipedia entry~\cite{HotaruWiki}). Circles representing
\emph{fireflies} (``hotaru'' in Japanese) are placed on a grid. The problem
asks to connect all the fireflies by drawing lines (representing \emph{beams} of
light) that follow a few rules, such as bending a prescribed number of times.

The main contribution of this paper is developing a physical ZKP using cards for
proving that one knows a solution to a Hotaru Beam puzzle. While the
connectivity constraint is a feature shared by several other types of puzzles
(e.g. Numberlink or Bridge) for which physical ZKPs already
exist~\cite{RuangwisesI21b,RuangwisesI21}, integrating the constraint on the
number of bends presents a unique challenge.
To address this, we introduce two innovative ideas. Firstly, we
propose a protocol for implementing 
beams while revealing the number of
bends. Secondly, we introduce a protocol for keeping track of
connections between fireflies throughout the process of implementing
beams. We believe these ideas offer valuable insights for
designing ZKPs for various other geometric objects.

Additionally, we prove the NP-completeness of Hotaru Beam. Our proof is
accomplished by a reduction from the planar monotone 3-SAT
problem~\cite{BergK12}.

The remainder of this paper is structured as follows:
Section~\ref{sec.rule} provides the formal definition of Hotaru
Beam. Section~\ref{sec.NP-completeness} demonstrates the NP-completeness
of Hotaru Beam. Section~\ref{sec.prelim} introduces fundamental concepts
regarding physical ZKPs, while Section~\ref{sec.protocol} details our
protocol designed for Hotaru Beam. 
Finally, Section~\ref{sec.conclusion} offers concluding remarks.

\section{The Rules of Hotaru Beam}
\label{sec.rule}

\begin{figure}[tb]
\centering
 \begin{tikzpicture}[scale=.7]
  \draw[backg] (0,0) grid (5,5);
  \node[hotaru] (a) at (1,1) {};
  \node[dot] at (a.east) {};
  \node[hotaru] (b) at (4,5) {2};
  \node[dot] at (b.south) {};
  \node[hotaru] (c) at (5,1) {1};
  \node[dot] at (c.north) {};
  \node[hotaru] (d) at (3,2) {};
  \node[dot] at (d.west) {};
  \node[hotaru] (e) at (1,4) {7};
  \node[dot] at (e.south) {};

\begin{scope}[xshift=250pt]
  \draw[backg] (0,0) grid (5,5);
  \node[hotaru] (a) at (1,1) {};
  \node[dot] at (a.east) {};
  \node[hotaru] (b) at (4,5) {2};
  \node[dot] at (b.south) {};
  \node[hotaru] (c) at (5,1) {1};
  \node[dot] at (c.north) {};
  \node[hotaru] (d) at (3,2) {};
  \node[dot] at (d.west) {};
  \node[hotaru] (e) at (1,4) {7};
  \node[dot] at (e.south) {};
 \draw [beam] (a) -- (3,1)--(d);
 \draw [beam] (b) -- (4,3)--(3,3)--(d);
 \draw [beam] (c) |- (b);
 \draw [beam] (d) -| (a);
 \draw [beam] (e) --(1,3) --(0,3)--(0,5)-|(2,4) -| (3,5)-- (b);
\end{scope}
 \end{tikzpicture}

\caption{An instance of Hotaru Beam (left) and a solution to it (right)}
\label{fig.instance}
\end{figure}
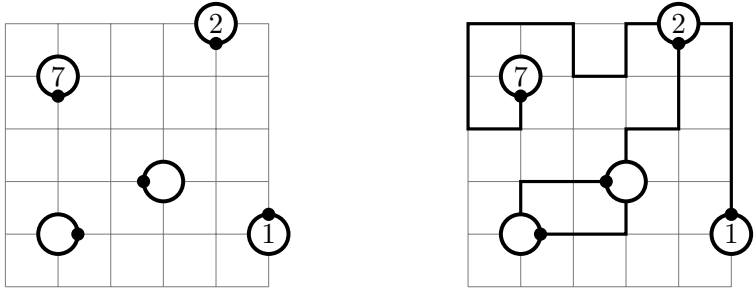

Hotaru Beam is played on a rectangular grid. Several circles, called fireflies,
are placed on the intersections of the grid. For each firefly, there is a dot
placed on one of the intersections of its boundary and grid lines, and there
also may be a number inside the circle. The left part of
Figure~\ref{fig.instance} illustrates an instance of Hotaru Beam.

The task is to draw a beam from each firefly. The beam starts from the dot,
follows the grid lines, and stops when it reaches a firefly's boundary (possibly
the same one as it started from). The beams drawn must satisfy the following
constraints.

\begin{itemize}
\item Beams do not intersect nor do they branch; therefore a beam cannot
\emph{end} at a firefly's dot since another beam (or the same beam) starts at that point.
\item The number inside a firefly prescribes the number of bends of its beam. We
call this the \emph{bending constraint}. If there is no number inside a firefly,
then the number of bends is not restricted.
\item All fireflies must end up connected into a single component by the beams.
We call this the \emph{connectivity constraint}. 
\end{itemize}

We refer to each individual line segment of a beam as a \emph{segment}.
A set of beams that satisfy these constraints is termed a \emph{solution}.
See the right panel of Figure~\ref{fig.instance} for an example of solutions.

\section{NP-completeness}
\label{sec.NP-completeness}

In this section, we prove the following theorem.

\begin{theorem}
It is NP-complete to decide whether an instance of Hotaru Beam admits
a solution.
\end{theorem}

Clearly, given a solution one can check its validity
in polynomial time, and hence the problem belongs to NP. We establish the
NP-completeness by a reduction from the planar monotone 3-SAT
problem~\cite{BergK12}.

\subsection{Planar Monotone 3-SAT}

The well-known 3-SAT problem is defined by a CNF formula $C_1 \land \dots \land C_m$
over a set of boolean variables $\{x_1, \dots, x_n\}$, where each clause $C_i$
is a disjunction of at most three literals (i.e. a variable or its negation). We
call a clause \emph{positive} if it consists only of positive literals;
similarly, we call a clause \emph{negative} if it consists only of negative
literals. A formula is called \emph{monotone} if each of its clause is positive
or negative.

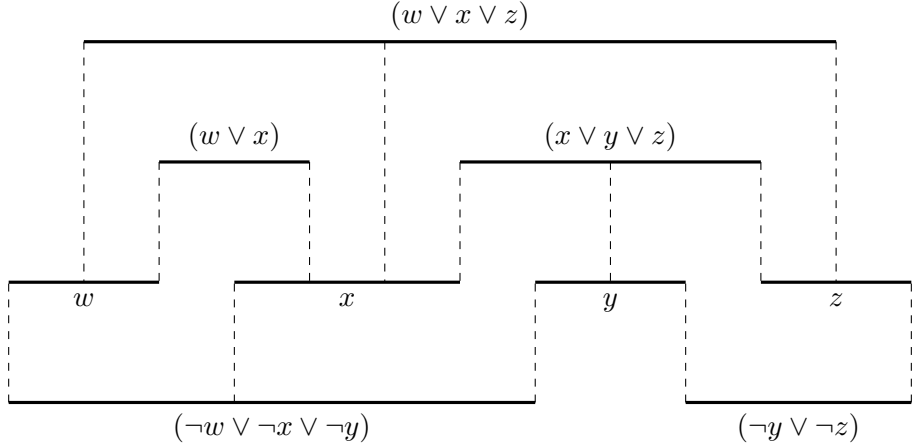
\begin{figure}[t]
\centering
 \begin{tikzpicture}[yscale=.8]

  \draw[line width=1.2pt] (0,0) -- node[below] {$w$} (2,0);
  \draw[line width=1.2pt] (3,0) -- node[below] {$x$} (6,0);
  \draw[line width=1.2pt] (7,0) -- node[below] {$y$} (9,0);
  \draw[line width=1.2pt] (10,0) -- node[below] {$z$} (12,0);

  \draw[line width=1.2pt] (1,4) -- node[above] {$(w \lor x \lor z)$} (11,4);
  \draw[line width=1.2pt] (2,2) -- node[above] {$(w \lor x)$} (4,2);
  \draw[line width=1.2pt] (6,2) -- node[above] {$(x \lor y \lor z)$} (10,2);
  \draw[line width=1.2pt] (0,-2) -- node[below] {$(\neg w \lor \neg x \lor \neg y)$} (7,-2);
  \draw[line width=1.2pt] (9,-2) -- node[below] {$(\neg y \lor \neg z)$} (12,-2);

  \draw[dashed] (0,-2) -- (0,0);
  \draw[dashed] (1,4) -- (1,0);
  \draw[dashed] (2,2) -- (2,0);
  \draw[dashed] (3,-2) -- (3,0);
  \draw[dashed] (4,2) -- (4,0);
  \draw[dashed] (5,4) -- (5,0);
  \draw[dashed] (6,2) -- (6,0);
  \draw[dashed] (7,-2) -- (7,0);
  \draw[dashed] (8,2) -- (8,0);
  \draw[dashed] (9,-2) -- (9,0);
  \draw[dashed] (10,2) -- (10,0);
  \draw[dashed] (11,4) -- (11,0);
  \draw[dashed] (12,-2) -- (12,0);

 \end{tikzpicture}

\caption{An instance of the planar monotone 3-SAT problem (the vertical segments
are drawn using dashed lines in order to improve clarity)}
\label{fig.planarSat}
\end{figure}

The planar monotone 3-SAT problem is a special case of the 3-SAT problem which
is given as a monotone formula represented by an embedding in the plane. More
precisely, the variables and clauses are each represented by a \emph{horizontal}
segment. The segments corresponding to the variables lie on the $x$-axis, while
the segments corresponding to the positive (negative) clauses lie above (below)
the $x$-axis. The occurrence of a variable in a clause is represented by a
\emph{vertical} segment connecting the two. Apart from this, the intersection of
segments is disallowed. (See Figure~\ref{fig.planarSat} for an example.)

The planar monotone 3-SAT problem is known to be NP-complete~\cite{BergK12}.

\subsection{Reduction}
We may without loss of generality assume that the endpoints of the segments have
integer coordinates within a range of a polynomial length (with respect to $m$
and $n$). This enables us to simply copy the placement of the segments when
constructing the grid of the Hotaru Beam instance.

\begin{figure}[t]
\centering
 \begin{tikzpicture}

  \draw[backg] (-0.5,-0.5) grid (10.5,4.5);

  \node[hotaru] (a) at (0,4) {0}; \node[dot] at (a.south) {};
  \node[hotaru] (b) at (10,4) {0}; \node[dot] at (b.west) {};
  \node[hotaru, fill=lightgray] (c) at (0,3) {1}; \node[dot] at (c.east) {};
  \node[hotaru] (d) at (2,2) {0}; \node[dot] at (d.east) {};
  \node[hotaru, fill=lightgray] (e) at (4,2) {1}; \node[dot] at (e.east) {};
  \node[hotaru] (f) at (6,1) {0}; \node[dot] at (f.east) {};
  \node[hotaru, fill=lightgray] (g) at (7,1) {1}; \node[dot] at (g.east) {};
  \node[hotaru] (h) at (9,0) {0}; \node[dot] at (h.east) {};
  \node[hotaru] (i) at (10,0) {0}; \node[dot] at (i.north) {};

  \draw[beam] (a) -- (c);
  \draw[beam] (d) -- (e);
  \draw[beam] (f) -- (g);
  \draw[beam] (h) -- (i);
  \draw[beam] (i) -- (b);
  \draw[beam] (b) -- (a);

  \draw[beam, dashed] (c) -- (2,3) -- (d);
  \draw[beam, dashed] (e) -- (5,2) -- (5,-0.5);
  \draw[beam, dashed] (g) -- (9,1) -- (h);

 \end{tikzpicture}

\caption{Gadget simulating a positive clause. The shaded fireflies represent the
literals; a possible set of beams shot from them is drawn using dashed lines. In
this example, only the second literal is going to be connected to the gadget
simulating its corresponding variable.}

\label{fig.gtClause}
\end{figure}
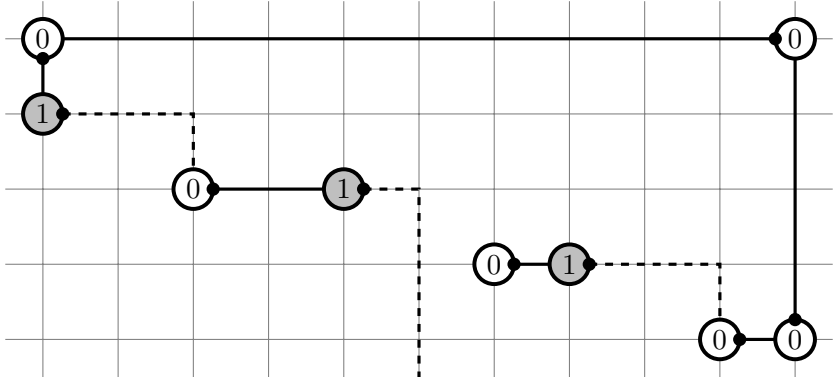

We replace variable segments with gadgets as shown in
Figure~\ref{fig.gtVariable}, positive clause segments with gadgets as shown in
Figure~\ref{fig.gtClause}, and negative clause segments with analogous gadgets
(flipped vertically). The number and positions of the shaded fireflies are
chosen so that they match the points of intersection with vertical segments.
Hence, for every vertical segment it is possible to shoot a beam from the
corresponding clause gadget arriving at the variable gadget. Note that a beam is
shot from every variable gadget and thus connects it to the next variable to the
right. In order to connect the rightmost variable to the leftmost one, we add a
chain of fireflies with bending number 0 such that it avoids the clause gadgets
(say, above all the positive clause gadgets).

If two points of intersection lie too close to each other, there may not be
enough space to unfold the gadgets. However, this issue can be easily avoided,
for example by assuming that the coordinates of the segment endpoints are all
multiples of a (sufficiently large) integer constant.

\begin{figure}[t]
\centering
 \begin{tikzpicture}[]

  \draw[backg] (-0.5,-2.5) grid (12.5,2.5);

  \node[hotaru] (a0) at (0,0) {0}; \node[dot] at (a0.east) {};
  \node[hotaru] (ap) at (0,+1) {0}; \node[dot] at (ap.south) {};
  \node[hotaru] (an) at (0,-1) {0}; \node[dot] at (an.north) {};
  \node[hotaru, fill=lightgray] (b1) at (1,0) {0}; \node[dot] at (b1.east) {};
  \node[hotaru, fill=lightgray] (b2) at (3,0) {0}; \node[dot] at (b2.east) {};
  \node[hotaru, fill=lightgray] (b3) at (6,0) {0}; \node[dot] at (b3.east) {};
  \node[hotaru] (c) at (7,0) {1}; \node[dot] at (c.east) {};
  \node[hotaru] (dp) at (8,+2) {0}; \node[dot] at (dp.east) {};
  \node[hotaru] (dn) at (8,-2) {0}; \node[dot] at (dn.east) {};
  \node[hotaru] (e) at (9,0) {2}; \node[dot] at (e.east) {};
  \node[hotaru] (f0) at (11,0) {0}; \node[dot] at (f0.east) {};
  \node[hotaru] (fp) at (11,+2) {0}; \node[dot] at (fp.south) {};
  \node[hotaru] (fn) at (11,-2) {0}; \node[dot] at (fn.north) {};

  \draw[beam] (ap) -- (a0);
  \draw[beam] (an) -- (a0);
  \draw[beam] (a0) -- (b1);
  \draw[beam] (b1) -- (b2);
  \draw[beam] (b2) -- (b3);
  \draw[beam] (b3) -- (c);
  \draw[beam] (dp) -- (fp);
  \draw[beam] (dn) -- (fn);
  \draw[beam] (fp) -- (f0);
  \draw[beam] (fn) -- (f0);
  \draw[beam] (f0) -- (12.5,0);

  \draw[beam, dashed] (c) -- (8,0) -- (dp);
  \draw[beam, dashed] (e) -- (10,0) -- (10,-1) -- (an);

 \end{tikzpicture}

\caption{Gadget simulating a variable. Beams from fireflies with bending number
0 (drawn using solid lines) are fixed in any solution. Note that there are only
two ways how to draw the beams from the fireflies with bending numbers 1 and 2.
The possibility drawn using the dashed lines corresponds to setting the variable
to true (the other case is symmetric). The shaded fireflies function as
potential entry points for beams from the clauses in which this variable
appears. In this case, only positive clauses (placed above) are going to be able
to connect to the variable. }

\label{fig.gtVariable}
\end{figure}
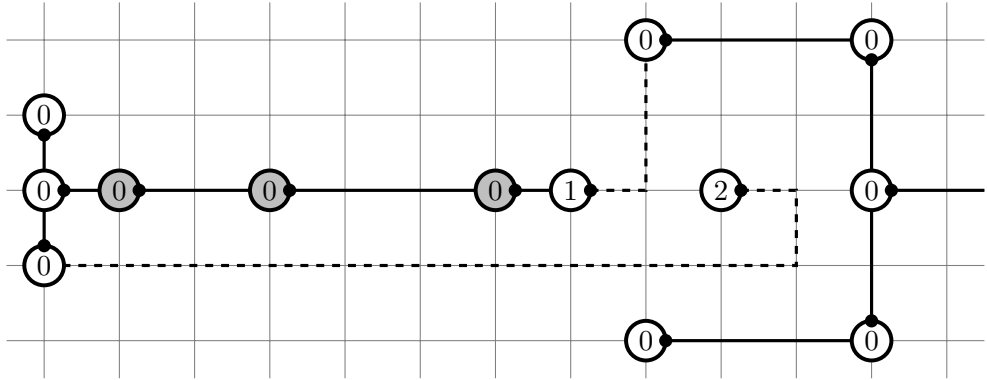

It is easy to see that the constructed Hotaru Beam instance is solvable if and
only if the given formula is satisfiable. First, consider a variable assignment
that satisfies the formula. For every variable gadget, we choose the way of
drawing beams depending on the value assigned to the variable, i.e., enabling
connection to the shaded fireflies from above (below) if the value is true
(false). For every shaded firefly (representing a literal) in clause gadgets, we
draw its beam leaving the gadget and arriving at the corresponding variable if
the assigned value satisfies the literal; otherwise we draw its beam so that it
connects to the next part of the same clause gadget. Note that the variable
gadgets form a loop and thus all fireflies in them are connected. Since the
variable assignment satisfies the formula, for every clause there is at least
one satisfied literal and hence all fireflies in the clause gadgets end up
connected to the fireflies in the variable gadgets.

For the other direction, consider a solution to the Hotaru Beam instance. We
define a variable assignment based on the choices of beam drawings inside the
variable gadgets (same as above). Since all the fireflies must be connected,
there is at least one beam leaving each clause gadget. By the construction, such
a beam may be shot only from a shaded firefly (representing a literal) and the
corresponding variable must be set so that the literal is satisfied. Therefore,
the assignment satisfies the formula.

\section{Preliminaries on Physical ZKP}
\label{sec.prelim}

\subsection{Definition of Physical ZKP}

A ZKP is a protocol for communication between two players, a prover (we will
call him Peter) and a verifier (we will call her Vera). In our case, both
players are presented with an instance $I$ of Hotaru Beam, while Peter is also
presented with a solution to $I$. His goal is to convince Vera that he knows a
solution without revealing any information about it.

The players exchange a series of messages as prescribed by the protocol. At the
end, Vera either accepts or rejects Peter's claim. The protocol must satisfy the
following three properties:
\begin{itemize}
\item \emph{Completeness}: If Peter knows a solution, then Vera accepts the
claim.
\item \emph{Soundness}: If Peter lacks knowledge of a solution, then Vera
rejects the claim.
\item \emph{Zero-knowledge}: Vera gains no additional information from the
communication beyond the fact that Peter is in possession of a solution.
\end{itemize}
In general, the players' behaviour can be randomized and the soundness condition
is relaxed so that Vera may accept a false claim with a small probability.
However, our protocol is deterministic with no allowance for error.

A common way to establish the zero-knowledge property is to provide an algorithm
(called a \emph{simulator}) that, without access to the solution, produces a
series of messages indistinguishable from those exchanged during a real run of
the protocol. In our case, the existence of a simulator will follow trivially
from the fact that every message directly depends only on the instance $I$.

As for a \emph{physical ZKP}, the protocol can be carried out by humans using
everyday objects. We present a card-based physical ZKP for Hotaru Beam. The
front of a card displays either a pictorial pattern (\club, \spade, \heart,
\diam) or a numerical value (\numcard{1}, \numcard{2}, $\ldots$), while the
backs of the cards are all indistinguishable from each other. Vera can see the
front of a card only if it is facing up. Peter, however, is allowed to freely
inspect any card, and thus we may as well assume that he can see the fronts of
all the cards. For clarity, we will represent face-down cards with \bclub,
\bspade, \bheart, \bdiam, \bnumcard{1}, \bnumcard{2}, $\ldots$.

The players manipulate the cards as specified by our protocol: They may
rearrange cards that are already laid out, turn a card over, discard a card or
introduce a new one. It is Vera's role to verify that the fronts of the cards
visible to her are showing the patterns/values as required by the protocol. For
example, when Peter is asked to discard a \heart card from a face-down card
pile, he will first allow Vera to confirm the front before actually removing the
card.


\subsection{Basic card-manipulating protocols}

In this subsection, we describe a few basic card-manipulating protocols that are
used in our ZKP as subroutines.

\subsubsection{Pile-shifting shuffle}

The \emph{pile-shifting shuffle} protocol (introduced by Nishimura
et~al.~\cite{NishimuraHMS18}) takes a sequence of face-down card piles $P_0,
P_1, \ldots, P_{k-1}$, each of the same size. It then applies a circular shift
by an amount $x$ chosen uniformly at random from $\{0, 1, \ldots, k-1\}$. 
In other words, the resulting sequence is $P_x, P_{(x+1) \bmod k}, \ldots, P_{(x+k-1) \bmod k}$, and
 any circular shift of the original sequence becomes equally
likely. The value of $x$ is not disclosed to the players.

In order to physically carry out this protocol, both players take turns to shift
the sequence by an arbitrary amount until nobody is realistically able to keep
track of the overall amount shifted. Shuffling is the only situation in our ZKP
where randomness (or rather loss of players' knowledge about the state of the
cards) occurs.


\subsubsection{Pile choosing}

The \emph{pile choosing} protocol (introduced by Dumas
et~al.~\cite{Dumas0MMSS19} in the form of a \emph{card choosing} protocol)
allows Peter to perform an operation on a card pile chosen from a sequence
without disclosing to Vera which pile was chosen. Given a sequence of face-down
card piles (each of the same size), the protocol proceeds as follows.

\begin{enumerate}
\item Peter marks his chosen pile by placing a \bspade card on it and a \bclub
card on each of the remaining piles. (Vera cannot see the fronts of the placed
cards.)
\item Vera marks the first pile in the sequence by placing a \heart card on it
and a \diam card on each of the remaining piles. These cards are then turned
face-down.
\item A pile-shifting shuffle is applied to the sequence. Peter identifies his
chosen pile by locating the \bspade card. He removes the top two cards (placed
there during the previous two steps), performs his intended operation on the
rest of the pile, and returns the two cards to the top. A pile-shifting shuffle
is applied once again to the sequence.
\item Vera turns the top card of every pile face-up, applies a circular shift so
that the pile marked by the \heart card returns to its original position (the
first pile in the sequence), and then removes the top two cards of every pile.
\end{enumerate}

Note that the zero-knowledge property is satisfied since the outcomes observable
by Vera (on which pile was the operation performed and the location of the
\heart card) are uniformly distributed.

\subsubsection{Reversible pile-shifting shuffle}

Extending the pile choosing protocol, this protocol allows Peter to perform
several operations on a number of chosen card piles from a sequence. Vera will
learn the relative positions of the chosen piles (i.e., the distances between
them along the cycle), but not their absolute position in the input sequence. In
other words, Peter can perform operations on a pile-shifting shuffled sequence
and then undo the shuffle (convincing Vera that the sequence is back to its
original order).

The implementation is actually the same as for the pile choosing protocol. Although it would be possible to always use this (more general) version,
invoking the special case (i.e., choosing a single pile rather than performing a
reversible shuffle) sometimes turns out to be more intuitive.

\subsubsection{Proving set membership}

This protocol can be used to persuade Vera that the front of a face-down card
$c$ belongs to a certain set $S$ without revealing the actual pattern/value.

A sequence of $|S|$ new cards with fronts $S$ is laid out face-up. After Vera
confirms their fronts, the cards are turned face-down and a pile-shifting
shuffle is applied. Peter exchanges card $c$ and the card in the sequence with
the same front as $c$. Another pile-shifting shuffle is applied to the sequence,
after which the cards are turned face-up. Finally, Vera confirms that the
sequence still consists of cards with fronts $S$.

\subsection{Representation of logical values}

We represent logical values by ordered pairs of cards: \club\heart corresponds
to \emph{true} and \heart\club corresponds to \emph{false}. Such a pair is
referred to as a \emph{logical card pair}. For simplicity, we denote a pair
representing true (false) by \bbinary{T} (\bbinary{F}). Although this may appear
as a single card, please note that it is actually an ordered pair.

We also introduce a protocol for Peter to replace two face-down logical card pairs $b_1$ and $b_2$ with their disjunction $b_1 \lor b_2$, without disclosing 
any information about the values to Vera.
The protocol proceeds as follows.

\begin{enumerate}
 \item Firstly, 
       a random shuffle is applied to $b_1$ and $b_2$ so 
       that the players cannot distinguish them. 
       Then, Peter examines the faces of the cards, and chooses one from $b_1$ and $b_2$. 
       If either $b_1$ or $b_2$ represents true, the Peter chooses it. 
       If both $b_1$ and $b_2$ represent true, or if they both
       represent false, Peter chooses one of them arbitrarily. 
       The chosen pair is placed in the first row.
       Throughout this process, the faces of the cards remained concealed, preventing Vera from determining which pair is
       chosen.

\item Another card pair \spade \diam is prepared and placed in the second row. Vera
      confirms the faces of the cards before they are turned
       face-down.

 \item The pile-shifting shuffle is applied, 
       considering the cards arranged vertically as a pile.

 \item The cards in the first row are flipped.
       Either player duplicates the same logical card pair from the first row and places it in the third row.
       In other words, the first row is replicated in the third row.
       Both players can confirm that the first and the third rows both represent the same logical value.

\item 
After flipping the cards in the first and the third rows back down, a pile-shifting shuffle is applied to all rows to shuffle the columns.

\item The cards in the second row are flipped, and
the piles are shifted so that the second row displays \spade \diam. 
At this stage, the first and third rows match the original first row.

\item Finally, the card pairs in the first and third rows are outputted.
\end{enumerate}

If both $b_1$ and $b_2$ are false, then the first row in Step 1 also
represents false. Consequently, both outputted card pairs are false in
this scenario. If $b_1$ or $b_2$ is true and the prover places it in
the first row in Step 1, then both outputted card pairs are true. In
these cases, the process effectively replaces the input logical card
pairs with card pairs representing their disjunction.

On the other hand, when exactly one of $b_1$ and $b_2$ is true, the
verifier cannot detect if the prover places the false pair on the first
row, which does not adhere to the specified behavior in Step 1,
resulting in the replacement of $b_1$ and $b_2$ with false card
pairs. However, this discrepancy does not pose a problem when used in
our ZKP for Hotaru beam because such behavior merely puts the prover at
a disadvantage.

The zero-knowledge of this protocol is immediate since the verifier's observation is independent from the prover's behavior.

\section{Physical ZKP for Hotaru beam}
\label{sec.protocol}

Throughout the rest of this paper, we denote by $n$ the number of fireflies, by
$h$ the height of the board (i.e. the number of grid points along the vertical
axis), and by $w$ the width of the board (i.e. the number of grid points along
the horizontal axis). We also arbitrarily fix an ordering of the fireflies.

In our protocol, Peter is going to use cards to encode his solution. He will
embed the beam starting from the $i$-th firefly sequentially for $i = 1, \ldots,
n$. The state after each embedding is captured using two structures, the board
and the connections table, which we describe in the following.

\subsection{Board}
\label{sec.board}

\begin{figure}[t]
\centering
 \begin{tikzpicture}[]
  \node[] at (0,2.4){\club \ \club \ \club \ \club \ \club \  \club \  \club \  \club};
  \node[] at (0,1.8){\club \ \heart \ \heart \heart \ \heart \  \numcard{5}  \heart  \club};
  \node[] at (0,1.2){\club \ \heart \ \numcard{1} \heart \ \heart \  \heart \  \heart  \club};
  \node[] at (0,.6){\club \ \heart \ \heart \ \heart \ \heart \  \heart \  \heart \  \club};
  \node[] at (0,0){\club \ \heart \  \heart \  \heart \numcard{3} \heart \  \heart \  \club};
  \node[] at (0,-.6){\club \ \heart \ \numcard{4} \heart \ \heart \  \heart \  \numcard{2}  \club};
  \node[] at (0,-1.2){\club \ \heart \ \heart \ \heart \ \heart \  \heart \  \heart \  \club};
  \node[] at (0,-1.8){\club \ \club \ \club \ \club \ \club \  \club \  \club \  \club};
 \end{tikzpicture}

 \caption{Card representation of the instance in Figure~\ref{fig.instance}}
 \label{fig.representation}
\end{figure}
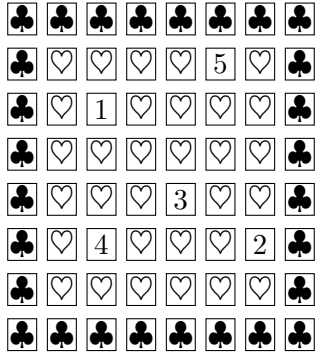

The board is initially represented by laying out cards as in
Figure~\ref{fig.representation}. Formally, we lay out $w \times h$ cards
corresponding to the grid points; the grid points containing fireflies are
represented with corresponding number cards (\numcard{$i$} for the $i$-th
firefly) while the remaining grid points are represented with \heart cards.
These $w \times h$ cards are then surrounded with a boundary made of \club
cards. We use \heart cards to express that a grid point is still available for a
beam to pass through; some of them will be replaced with \club cards during the
protocol as we will sequentially embed the beams of the solution.

We will refer to these cards simply as \emph{the board}. After Vera confirms the
initial arrangement of the board, all the cards are turned face-down.

\subsection{Connections table}
\label{sec.connectionstable}

In addition to the board, our protocol maintains a set of cards encoding
connections between the fireflies which we call the \emph{connections table}
(refer to Figure~\ref{fig.connectionstable}). It consists of $n$ columns, one
for each firefly. The $i$-th column consists of an \numcard{$i$} card followed
by $n$ logical card pairs. Initially, we use a \bbinary{T} for the $i$-th pair
and a \bbinary{F} for the remaining pairs.

After Vera confirms the initial arrangement of the connections table, all the
cards are turned face-down.

Let $G$ be an undirected graph on $n$ vertices (representing the fireflies) with
edges corresponding to the beams that were embedded up to that point. The
following invariant will hold throughout our protocol: If the $j$-th logical
card pair in the $i$-th column of the connections table is set to \bbinary{T},
then Vera can be sure that fireflies $i$ and $j$ belong to the same connected
component of $G$.

\begin{figure}[t]
\centering
 \begin{tikzpicture}[xscale=.7]
  \node[] at (0,2.4){\numcard{1} \  \numcard{2} \  \numcard{3}  \  \numcard{4} \  \numcard{5}};
  \node[] at (0,1.8){\bbinary{T} \  \bbinary{F} \  \bbinary{F} \  \bbinary{F} \  \bbinary{F}};
  \node[] at (0,1.2){\bbinary{F} \  \bbinary{T} \  \bbinary{F} \  \bbinary{F} \  \bbinary{F}};
  \node[] at (0,.6){\bbinary{F} \  \bbinary{F} \  \bbinary{T} \  \bbinary{F} \  \bbinary{F}};
  \node[] at (0,0){\bbinary{F} \  \bbinary{F} \  \bbinary{F} \  \bbinary{T} \  \bbinary{F}};
  \node[] at (0,-.6){\bbinary{F} \  \bbinary{F} \  \bbinary{F} \  \bbinary{F} \  \bbinary{T}};
 \end{tikzpicture}
 \caption{The initial state of the connections table in case of 5 fireflies}
 \label{fig.connectionstable}
\end{figure}

\subsection{Embedding a beam segment}

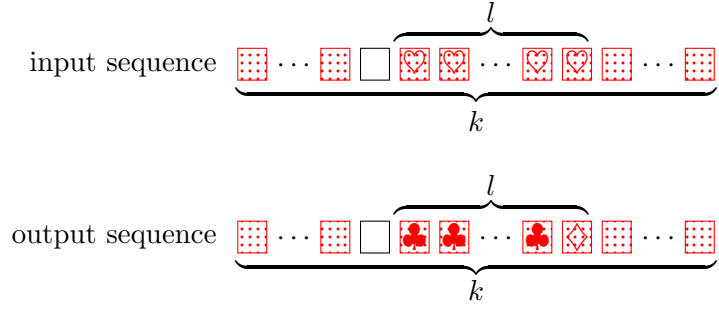
\begin{figure}[t]
\centering

 \begin{tikzpicture}
  \node at (0,-.75){$k$};
  \node at (0,-.4) {$\underbrace{\hspace*{16.6em}}$};
  \node at (.2,.7){$l$};
  \node at (.2,.4) {$\overbrace{\hspace*{6.7em}}$};

  \node[label=left:input sequence] at (0,0){\bblank $\cdots$ \bblank \blank \bheart \bheart $\cdots$ \bheart \bheart \bblank $\cdots$ \bblank};

  \node at (0,-3){$k$};
  \node at (0,-2.7) {$\underbrace{\hspace*{16.6em}}$};
  \node at (.2,-1.6){$l$};
  \node at (.2,-1.9) {$\overbrace{\hspace*{6.7em}}$};

  \node[label=left:output sequence] at (0,-2.3){\bblank $\cdots$ \bblank \blank \bclub \bclub $\cdots$ \bclub \bdiam \bblank $\cdots$ \bblank };
 \end{tikzpicture}
\caption{Input and output sequences of the segment embedding protocol}
\label{fig.reversi}
\end{figure}

In this subsection, we describe the \emph{segment embedding} protocol, which
Peter will use to update the board by embedding a part (a beam segment) of his
solution. The protocol takes as the input a sequence of $k$ cards (a row or a
column of the board). One of them is turned face-up (we will refer to it as the
\emph{start} card), while the rest is face-down. Peter can choose a value $l$
and replace $l$ consecutive cards adjacent to the start card, assuming they are
all \heart cards. See Figure~\ref{fig.reversi} for an example.

We use several variations of the protocol which differ in restrictions on the
replacement cards and what information is revealed to Vera. 
First, we describe
the most basic version; the implementation details of the other variants are
discussed at the end of this section.

In the basic setting, Peter is required to choose a value $l \geq 1$. The exact
value is not revealed to Vera, but she can be sure it satisfies the requirement.
(In some variants, Peter is also allowed to choose $l = 0$ without Vera being
able to tell the difference.) Peter can choose to replace cards to the left or
to the right of the start; however, this choice is revealed to Vera. (In some
variants, this choice is kept hidden from her.) Beginning from the card adjacent
to the start, Peter replaces the first $l-1$ \heart cards with \club cards and
the last \heart card with a \diam card. (In some variants, the last \heart card
is also replaced with a \club card. In that setting, it is required that a
number card follows after the last \heart card. Vera learns that it is the case,
but gains no information about the value of the number card.)

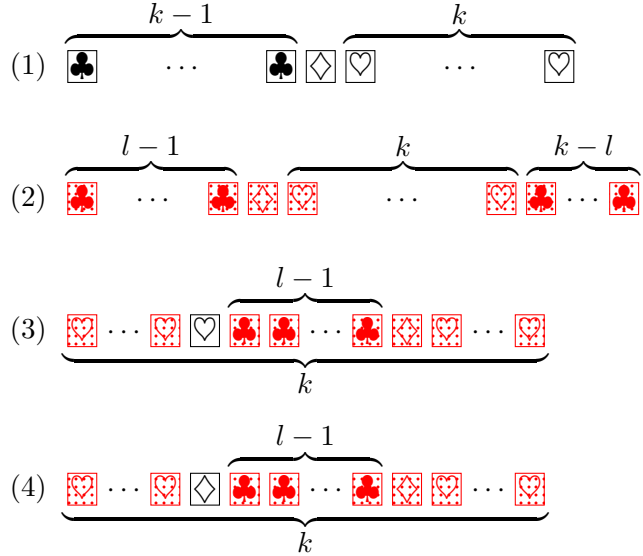
\begin{figure}[t]
\centering
 \begin{tikzpicture}
  \node at (1.65,.7){$k-1$};
  \node at (1.65,.4) {$\overbrace{\hspace*{8em}}$};
  \node at (5.35,.7){$k$};
  \node at (5.35,.4) {$\overbrace{\hspace*{8em}}$};
  \node[label=left:(1),anchor=west] at (0,0){\club \hspace*{2em}$\cdots$\hspace*{2em} \club \diam  \heart \hspace*{2em}$\cdots$\hspace*{2em} \heart};

\begin{scope}[yshift=-50pt]
  \node at (1.25,.7){$l-1$};
  \node at (1.25,.4) {$\overbrace{\hspace*{5.9em}}$};
  \node at (4.6,.7){$k$};
  \node at (4.6,.4) {$\overbrace{\hspace*{7.95em}}$};
  \node at (7,.7){$k-l$};
  \node at (7,.4) {$\overbrace{\hspace*{3.9em}}$};
  \node[label=left:(2),anchor=west] at (0,0){\bclub \hspace*{1em}$\cdots$\hspace*{1em} \bclub \bdiam  \bheart \hspace*{2em}$\cdots$\hspace*{2em} \bheart \bclub $\cdots$ \bclub};
\end{scope}

\begin{scope}[yshift=-100pt]
  \node at (3.3,.7){$l-1$};
  \node at (3.3,.4) {$\overbrace{\hspace*{5.3em}}$};
  \node[label=left:(3),anchor=west] at (0,0){\bheart $\cdots$ \bheart \heart \bclub \bclub $\cdots$ \bclub \bdiam  \bheart $\cdots$ \bheart};
  \node at (3.3,-.7){$k$};
  \node at (3.3,-.4) {$\underbrace{\hspace*{16.8em}}$};
\end{scope}

\begin{scope}[yshift=-160pt]
  \node at (3.3,.7){$l-1$};
  \node at (3.3,.4) {$\overbrace{\hspace*{5.3em}}$};
  \node[label=left:(4),anchor=west] at (0,0){\bheart $\cdots$ \bheart \diam \bclub \bclub $\cdots$ \bclub \bdiam  \bheart $\cdots$ \bheart};
  \node at (3.3,-.7){$k$};
  \node at (3.3,-.4) {$\underbrace{\hspace*{16.8em}}$};
\end{scope}

 \end{tikzpicture}
\caption{A mask sequence being constructed step-by-step}
\label{fig.reversi_initial}
\end{figure}

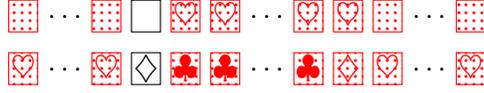
\begin{figure}[ht]
\centering
 \begin{tikzpicture}
  \node[] at (0,0){\bblank $\cdots$ \bblank \blank \bheart \bheart $\cdots$ \bheart \bheart \bblank $\cdots$ \bblank};
  \node[] at (0,-.7){\bheart $\cdots$ \bheart \diam \bclub \bclub $\cdots$ \bclub \bdiam \bheart $\cdots$ \bheart};
 \end{tikzpicture}
\caption{The arrangement of the cards after constructing the mask sequence}
\label{fig.reversi_step3}
\end{figure}

Suppose that Peter chose a value $l \geq 1$ and wishes to replace cards to the
right of the start (the case of replacing cards to the left is symmetric).
First, we construct an auxiliary sequence of $k$ cards called the \emph{mask}
sequence (see the bottom row in Figure~\ref{fig.reversi_step3}). This is done as
follows:

\begin{enumerate}
\item The players lay out a sequence of new cards as in the first row of
Figure~\ref{fig.reversi_initial}. After Vera confirms their fronts, these cards
are all turned face-down.
\item The sequence is shuffled using the pile-shifting shuffle protocol. Peter
then applies a further circular shift in order to bring the \diam card to the
$l$-th position from the left, as in the second row of Figure~\ref{fig.reversi_initial}.
\item The cards to the right of the $k$-th card are discarded, after which a
pile-shifting shuffle is applied to the remaining sequence. Peter turns the
\diam card face-up to show Vera that it is still present in the sequence. Then
he turns it back face-down, applies another pile-shifting shuffle, and turns the
first \heart card to the left of the \diam card face-up, as in the third row of Figure~\ref{fig.reversi_initial}.
\item Peter replaces the face-up \heart card with a \diam card. He then proves
to Vera (using the proving set membership protocol) that the next card to the
right is a \club card or a \diam card.
\end{enumerate}

Note that Vera can be sure that the resulting mask sequence looks as in the last
row of Figure~\ref{fig.reversi_initial} for some $l \geq 1$, but she does not
learn anything else about the value of $l$.

The mask sequence is circularly shifted and placed beneath the input sequence so
that the face-up \diam card aligns with the start card (see
Figure~\ref{fig.reversi_step3}). Consider any face-down column (a pair of
vertically aligned cards) and note that it must contain at least one \heart
card. For every such column, Peter shuffles the pair and removes from it a
\heart card. He allows Vera to verify the front of the removed card and then
discards it. The remaining card is returned to the top row. Finally, the \diam
card aligned with the start card is also discarded. The result looks as in
Figure~\ref{fig.reversi}.

Finally, we discuss a few other variants of embedding a beam segment that are
used in our protocol.
\begin{itemize}
\item \emph{Hiding the direction of the replacement} \\
Peter prepares two versions of the mask sequence: one for replacing cards to the
left of the start and one for replacing cards to the right. He then secretly
chooses the required one (applying a shuffle) and uses it for the replacement.
The other version is discarded.
\item \emph{Allowing $l \geq 0$} \\
If $l = 0$, in reality no replacement occurs. This can be simulated with a mask
sequence that consists of $k-1$ \heart cards and a single \diam cards. To make
this case indistinguishable, Peter also prepares an unnecessary mask sequence
for some $l' \geq 1$ and secretly chooses the required one.
\item \emph{Embedding the final segment} \\
The final segment of a beam ends at a firefly, and thus the $l$-th card is a
number card rather than a \heart card. Peter will not reveal and discard a
\heart card from that column; instead, he will put the face-down number card
away. As described in the following section, Peter can convince Vera that this
is really a number card without disclosing its value.
\end{itemize}

\subsection{Embedding a beam}

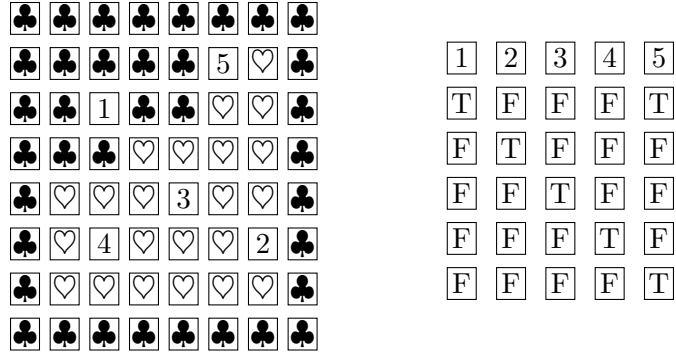
\begin{figure}[t]
\centering
 \begin{tikzpicture}[]
  \node[] at (0,2.4){\club \ \club \ \club \ \club \ \club \  \club \  \club \  \club};
  \node[] at (0,1.8){\club \ \club \ \club \ \club \ \club \  \numcard{5}  \heart  \club};
  \node[] at (0,1.2){\club \ \club \ \numcard{1} \club \ \club \  \heart \  \heart  \club};
  \node[] at (0,.6){\club \ \club \ \club \ \heart \ \heart \  \heart \  \heart \  \club};
  \node[] at (0,0){\club \ \heart \  \heart \  \heart \numcard{3} \heart \  \heart \  \club};
  \node[] at (0,-.6){\club \ \heart \ \numcard{4} \heart \ \heart \  \heart \  \numcard{2}  \club};
  \node[] at (0,-1.2){\club \ \heart \ \heart \ \heart \ \heart \  \heart \  \heart \  \club};
  \node[] at (0,-1.8){\club \ \club \ \club \ \club \ \club \  \club \  \club \  \club};

 \begin{scope}[xshift=150pt,yshift=-15pt]
  \node[] at (0,2.4){\numcard{1} \  \numcard{2} \  \numcard{3}  \  \numcard{4} \  \numcard{5}};
  \node[] at (0,1.8){\bbinary{T} \  \bbinary{F} \  \bbinary{F} \  \bbinary{F} \  \bbinary{T}};
  \node[] at (0,1.2){\bbinary{F} \  \bbinary{T} \  \bbinary{F} \  \bbinary{F} \  \bbinary{F}};
  \node[] at (0,.6){\bbinary{F} \  \bbinary{F} \  \bbinary{T} \  \bbinary{F} \  \bbinary{F}};
  \node[] at (0,0){\bbinary{F} \  \bbinary{F} \  \bbinary{F} \  \bbinary{T} \  \bbinary{F}};
  \node[] at (0,-.6){\bbinary{F} \  \bbinary{F} \  \bbinary{F} \  \bbinary{F} \  \bbinary{T}};
 \end{scope}
 \end{tikzpicture}

 \caption{Embedding the beam from the 1st firefly to the 5th firefly into the
 board (left) and the connections table (right)}
 \label{fig.solution}
\end{figure}

Suppose that Peter wants to embed a beam starting from the $s$-th firefly and
ending at the $t$-th firefly. This involves modifying both the board (replacing
the \heart cards lying on the grid points passed by the beam with \club cards)
and the connections table (setting the $s$-th logical card pair in the $t$-th
column to \bbinary{T}). See Figure~\ref{fig.solution} for an example.

If the beam is forbidden to bend at all (i.e., the $s$-th firefly has a $0$
inside or the grid point next to its dot contains another firefly), then it is
drawn the same way in any solution. Hence, there is no need to hide its drawing
and Peter may thus make the necessary changes so that Vera sees them. In the
rest, we will assume that this special case does not apply.

First, Peter creates a copy of the logical card pairs in the $s$-th column of
the connections table (without revealing to Vera the actual values). We denote
this copied sequence by $C$.

Embedding a beam can be done by breaking it into straight parts (segments) and
iteratively applying the segment embedding protocol. Let us first consider the
case when the beam to embed has a prescribed number of bends $b \geq 1$. We
split the beam at the bends and thus obtain $b+1$ segments.

Suppose that the direction of the first segment is horizontal, the other case is
analogous (note that the direction is determined by the position of the dot, and
hence is known to Vera). Peter is going to embed it by applying the segment
embedding protocol to the row of the board which contains the $s$-th firefly. He
turns the \numcard{$s$} card face-up to designate it as the start card, sets $l$
to equal the length of the segment ($l \geq 1$), and publicly chooses the
direction of replacement (left or right) following the position of the dot.
Applying the protocol, he replaces the first $l-1$ \heart cards with \club cards
and the last \heart card with a \diam card. This \diam card serves as the start
card when embedding the next segment.

The direction of the remaining segments alternates between vertical and
horizontal. In case of a vertical (horizontal) segment, Peter applies the pile
choosing protocol to pick up the column (row) $X$ of the board that contains the
\diam card without revealing its position to Vera. He then applies the
reversible pile-shifting shuffle to $X$ and turns the \diam card face-up. He
sets $l$ to equal the length of the segment to embed ($l \geq 1$), secretly
chooses the direction of replacement, and applies the appropriate variant of the
segment embedding protocol. Then he undoes the shuffle performed on $X$ and
returns it to the board.

When embedding the last segment, Peter sets $l$ to equal one less than its
length ($l \geq 0$) and replaces \emph{all} the \heart cards with \club cards.
Let us denote by $f$ the card that follows after the last replaced card (it is
in fact \numcard{$t$}). Peter needs to convince Vera that $f$ is a number card
without revealing its value or position. This can be accomplished similarly as
in the proving set membership protocol: Peter picks up the $t$-th column of the
connections table (using the pile choosing protocol) and exchanges the
\numcard{$t$} in its first row with $f$. For every logical card pair $p$ in the
column, he applies the logical-values copy protocol to $p$ and its corresponding
logical card pair $p'$ in $C$. If $p = \numcard{F}$, he replaces $p$ with $p'$;
otherwise he replaces $p'$ with $p$. In other words, after the operation, $p$
equals the disjunction of the two original logical values. This ensures that the
$s$-th logical card pair in the $t$-th column is set to \numcard{T}, since the
$s$-th logical card pair in the $s$-th column was initialized to \numcard{T}.
After returning the $t$-th column to the connection table, Peter lets Vera
confirm that the top row still contains cards \numcard{1}, \dots, \numcard{$n$}.
Vera can now be sure that $f$ is indeed a number card and that Peter performed
the logical-values copy protocol on the appropriate column. Although she did not
learn anything about the values and how they were replaced, the invariant of the
connections table is maintained since the fireflies involved ($s$ and $t$) have
just been connected by a beam.

If the number of times the beam bends is not prescribed by the instance, it must
be hidden from Vera. Peter accomplishes that by adding \emph{dummy} segments of
length $0$ so that the number of segments becomes $w \cdot h$. In this
case, he utilizes variants of the segment embedding protocol that allow $l = 0$.
Note that if the beam does not bend at all, it still must be split into at
least two (non-dummy) segments, since we treat the first and the last segments
in a special way.

\subsection{Overall description of the protocol}
\label{sec.overall}

Initially, the board and the connections table cards are arranged as described in
Sections~\ref{sec.board} and~\ref{sec.connectionstable}, 
Vera confirms their fronts, and all the cards are turned face-down.
Then, Peter sequentially
embeds the beam starting from the $s$-th firefly for $s = 1, \ldots, n$.
Finally, he ensures that all logical card pairs in the connections table are
\numcard{T} by repeatedly performing the following operation.

Suppose that for two columns of the connections table (say, $i$ and $j$), there
is a row (say, $k$) in which both have a \numcard{T} logical card pair. By the
invariant, this implies that fireflies $i$ and $j$ belong to the same connected
component of $G$, and hence Peter may replace the two columns with their
row-wise disjunction in the same manner as when embedding the last segment of a
beam.

The columns involved in the operation are picked up using the pile choosing
protocol. Since graph $G$ is connected, all logical card pairs in the
connections table can be made into \numcard{T} by applying this operation at
most $n^2$ times. (Peter will always apply exactly $n^2$ operations to
hide the necessary number.) Finally, Vera turns the cards in the connections
table face-up and verifies their fronts. By the invariant, this convinces her
that the connectivity constraint is satisfied.

 \section{Conclusion}
 \label{sec.conclusion}

 We established the NP-completeness of Hotaru Beam and provided a physical ZKP
 for it. Our ZKP includes two novel ideas: the segment embedding protocol and the
 connections table. We believe that these are potentially useful in other
 protocols, particularly for puzzles on grid boards and for proving the existence
 of specific graphs.

\section*{Acknowledgements}
This work was supported by JSPS KAKENHI Grant Numbers JP20H05965, JP23K21646 and JP21K11759.

\end{document}